\documentclass{aa}  
\bibpunct{(}{)}{;}{a}{}{,} 
\usepackage{txfonts}

\usepackage{amssymb,amsmath,times,natbib}
\usepackage{graphicx}
\def\gsim{ \lower .75ex \hbox{$\sim$} \llap{\raise .27ex \hbox{$>$}} }
\def\lsim{ \lower .75ex\hbox{$\sim$} \llap{\raise .27ex \hbox{$<$}} }

\def\beq{\begin{equation}}
\def\eeq{\end{equation}}

\def\sw{{\it Swift}}
\def\fe{{\it Fermi}}
\def\ba{BATSE}

\def\ep{$E_{\rm p}$}

\def\eiso{$E_{\rm iso}$}
\def\ama{$E_{\rm p}-E_{\rm iso}$}

\def\th{$\theta_{\rm jet}$}

\def\thv{$\theta_{\rm view}$}

\def\G{$\Gamma_{0}$}

\begin{document}

\title{Unveiling the population of orphan Gamma Ray Bursts}

\author{
G. Ghirlanda
\inst{1}
\thanks{E--mail:giancarlo.ghirlanda@brera.inaf.it}, 
\and
R. Salvaterra\inst{2} \and S. Campana\inst{1} \and S. D. Vergani\inst{3,1} \and J. Japelj\inst{4} \and M. G. Bernardini\inst{1}  \and \\ D. Burlon\inst{5,6}  \and  P. D'Avanzo\inst{1} \and  A. Melandri\inst{1} \and  A. Gomboc\inst{4} \and  F. Nappo\inst{7,1} \and R. Paladini\inst{8} \and \\ A. Pescalli\inst{7,1} \and  O. S. Salafia\inst{9,1} \and  G. Tagliaferri\inst{1}
}
\institute{$^{1}$INAF -- Osservatorio Astronomico di Brera, via E. Bianchi 46, I-23807 Merate, Italy.\\
$^{2}$INAF -- IASF Milano, via E. Bassini 15, I-20133 Milano, Italy. \\
$^{3}$ GEPI, Observatoire de Paris, CNRS, Univ. Paris Diderot, 5 place Jules Janssen, 92190, Meudon, France.\\
$^{4}$Faculty of Mathematics and Physics, University of Ljubljana, Jadranska 19, 1000 Ljubljana, Slovenia.\\
$^{5}$Sydney Institute for Astronomy, School of Physics, The University of Sydney, NSW 2006, Australia.\\
$^{6}$ARC Centre of Excellence for All-sky Astrophysics (CAASTRO).\\
$^{7}$Universit\'a degli Studi dell'Insubria, via Valleggio 11, I-22100 Como, Italy.\\
$^{8}$NASA Herschel Science Center, California Institute of Technology, 1200, East California Boulevard, Pasadena, CA 91125, USA.\\
$^{9}$Dipartimento di Fisica G. Occhialini, Universit\'a di Milano Bicocca, Piazza della Scienza 3, I-20126 Milano, Italy.\\
}

\date{}


\abstract{
Gamma Ray Bursts are detectable in  the $\gamma$--ray band if their jets are oriented towards the observer. 
However, for each GRB with a typical \th, there should be $\sim 2/\theta_{\rm jet}^{2}$ bursts whose emission cone is oriented elsewhere in space. These off--axis bursts can be eventually detected  when, due to the deceleration of their relativistic jets, the beaming angle becomes comparable to the viewing angle. Orphan Afterglows (OA) should outnumber the current population of bursts detected in the $\gamma$--ray band even if they have not been conclusively observed so far at any frequency. We compute the expected flux of the population of orphan afterglows in the mm, optical and X-ray bands through a population synthesis code of GRBs and the standard afterglow emission model. We estimate the detection rate of OA by on--going and forthcoming surveys. The average duration of OA as transients above a given  limiting flux is  derived and described with analytical expressions: in general OA should appear as daily transients in optical surveys and as monthly/yearly transients in the mm/radio band. We find that $\sim$ 2 OA yr$^{-1}$ could already be detected by {\it Gaia} and up to 20 OA yr$^{-1}$ could be observed by the ZTF survey. A larger number of 50 OA yr$^{-1}$ should be detected by LSST in the optical band. For the X--ray band, $\sim$ 26 OA yr$^{-1}$ could be detected by the eROSITA. For the large population of OA detectable by LSST, the X--ray and optical follow up of the light curve (for the brightest cases) and/or the extensive follow up of their emission in the mm and radio band could be the key to disentangle their GRB nature from other extragalactic transients of comparable flux density. 
}

\maketitle

\section{Introduction}
Gamma Ray Bursts (GRBs) are cosmological sources which signpost the birth of stellar mass black holes or the most intense magnetic fields harbored by compact objects (magnetars). They are detected as short, highly variable, flashes of $\gamma$--rays (prompt emission) with  
duration that is typically in the range 0.1 to 1000 seconds 
followed by a smoothly decaying long lived emission at X--ray, optical, radio frequencies (the afterglow - Costa et al. \citeyear{Costa:1997la}; van Paradijs et al. \citeyear{van-Paradijs:1997it}). 
In the standard fireball model the prompt emission is interpreted as due to internal dissipation (either through relativistic shocks or magnetic reconnection) while the afterglow is produced by the deceleration of the relativistic outflow by the circum burst interstellar medium. 

Theoretical arguments \citep[e.g.][]{Sari:1999gd} and direct observational evidences \citep[e.g.][]{Molinari:2007ph} suggest that the outflow of GRBs is relativistic with typical bulk Lorentz factors \G$\sim 10^{2-3}$. The most luminous and energetic GRBs seem to have larger \G\ \citep[e.g.][]{Liang:2010cs,Ghirlanda:2012pi}. Another key property of GRBs is the presence of a jet. Invoked to reduce the otherwise huge isotropic equivalent energies by a factor proportional to $\theta^{2}_{\rm jet}$ (i.e. the jet half opening angle), jet angles have been estimated in a few tens of GRBs from the steepening of the optical and/or X--ray light curve a few days after the prompt emission. This steepening is interpreted as the time when, due to the deceleration of the outflow, the {\it relativistic beaming} $\propto 1/\Gamma$ equals the {\it geometric collimation} \th. Modelling of the outflow dynamics \citep{Blandford:1976jt} allows us to infer the GRB jet opening angle \th\ \citep{Rhoads:1999hb}. The collimation corrected energies  were found to be distributed around $10^{51}$ erg \citep{Frail:2001sf} with a smaller dispersion with respect to that of the isotropic equivalent energies \citep[but see][]{Gehrels:2009fk}.

The satellites/instruments deputed to the detection of GRBs observed hundreds/thousands of GRBs at an average rate of $\sim$ 0.3 day$^{-1}$.  However, if GRBs have a jet with the energy and bulk Lorentz factors which are constant within the jet as a function of the angle from its axis and \G\ (pointed radially as the expansion of the outflow within the jet) is relatively large (typically a few hundreds), we can detect only those bursts whose jet is pointing at the Earth. Indeed, since the highly relativistic motion results in a strong forward beaming of the emitted radiation, the flux directed at the Earth is dramatically reduced when \thv$>$\th\ (where \thv\ is the viewing angle between the jet axis and the line of sight). These events (which are the most numerous due to the jet orientation probability being $\propto \sin \theta_{\rm view}$) go undetected as prompt $\gamma$--ray bursts.  However, during the afterglow the bulk Lorentz factor decreases with time (as the outflow is decelerated by the external medium). There will be a characteristic timescale when the {\it relativistic beaming} $\propto 1/\Gamma$ equals the {\it observer viewing angle} \thv\ and the (afterglow) radiation can be seen. These events, missing the prompt emission but detected as afterglows, are called {\it orphan afterglows} (OA - hereafter) and, for typical opening angles of GRBs of a few degrees, e.g. \th$\sim$0.1 rad, they should outnumber the population of GRBs (by a factor $\propto(1-\cos\theta_{\rm jet})^{-1}\sim200$).

Therefore, OA should be detected as transients but their association with GRBs is made difficult due to the lack of a prompt high--energy emission. Despite specific studies have been searching for OA in X--ray surveys \citep{Grindlay:1999yf,Greiner:2000zp} in optical surveys \citep{vreeswijk:2002qy,Rau:2006bf,Malacrino:2007ud,Rau:2007kk} and in the radio band \citep{Levinson:2002lr,Gal-Yam:2006wq,Bannister:2011ao,Bell:2011zm,Bower:2011ix,Croft:2010rp,Frail:2012if,Carilli:2003ef,Matsumura:2009fe,Lazio:2010cl}, no OA has been conclusively  detected so far. 

Non detections of OA are in agreement with current theoretical predictions \citep{Totani:2002hb,Nakar:2002vf,Zou:2007uk,Rossi:2008om,Metzger:2015eq}. However, these works either extrapolated the properties of a few known GRB afterglows to the orphans \citep[e.g.][]{Totani:2002hb} or assumed basic prescriptions for the known GRB population properties or for the afterglow emission model. We have recently developed a population synthesis code \citep{Ghirlanda:2013fq} which, coupled with the most detailed model for the afterglow emission \citep{van-Eerten:2012oj}, allowed us to predict the properties of the population of OA \citep{Ghirlanda:2014wt} reproducing a large set of observed properties of the population of the ``Earth--pointed'' GRBs. We have, so far, considered the radio band predicting that the Square Kilometer Array (SKA), reaching the $\mu$Jy flux limit, could see up to $\sim$0.2--1.5 OA deg$^{-2}$ yr$^{-1}$ \citep{Ghirlanda:2014wt}. Alternatively, the non detection of OA could be due to the structure of the jet \citep{Rossi:2008om,Salafia:2015oq}. 

We are entering the era of large synoptic surveys which will monitor large portions of the (if not the whole) sky with unprecedented sensitivities. OA are potentially in the list of transients that these surveys will detect, but specific predictions on the rate depend on the true rate of the population of OA (and also their duration) and on the survey characteristics (area of the sky covered, timescales, limiting flux). Here we derive the flux distribution of OA in the X--ray, optical and mm band (\S 3) based on our recent population synthesis code (\S 2). We also study the average duration of the population of OA as a function of the survey limiting flux (\S 3).  We compare with current limits of OA in these bands and make predictions for on--going and forthcoming surveys (\S 4). Standard cosmological parameters ($h=\Omega_{\Lambda}=0.7$) for a flat Universe are adopted throughout the paper. 

\section{Orphan Afterglow emission}
\begin{center}
\begin{figure}
\hskip -0.3truecm
\includegraphics[width=8.5cm,trim=20 10 20 20,clip=true]{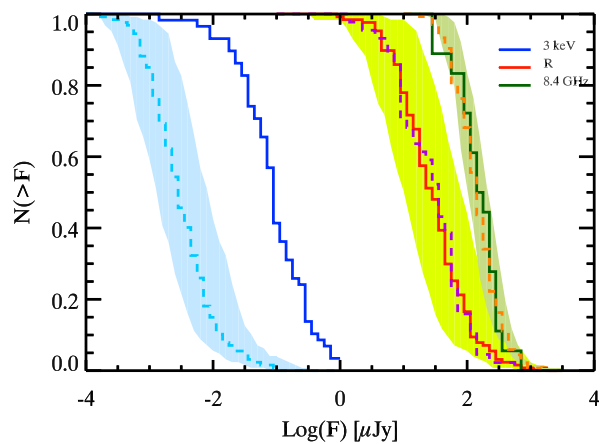}
\caption{Flux density cumulative distributions in the optical ($R$ - solid red line) X--ray (at 3 keV - solid blue line) computed at 11 h after the start of the GRB emission for the BAT6 \sw\ complete sample (adapted from Melandri et al. \citeyear{Melandri:2014lk} and D'Avanzo et al. \citeyear{DAvanzo:2012wj} for the optical and X--ray band, respectively). The radio (at 8.4 GHz - solid green line) is from \citet{Ghirlanda:2013bq}, also for the BAT6 sample. The results of the population synthesis code with $p=2.3, \epsilon_e=0.02$ and $\epsilon_B=0.008$ are shown with the dashed lines. The shaded regions represent, for each band, the results obtained with $(p,\epsilon_e,\epsilon_B)=(2.3,0.01,0.001)$ for the lower boundary and $(p,\epsilon_e,\epsilon_B)=(2.3,0.05,0.01)$ for the upper boundary. The X--ray  and $R$ band fluxes of the \sw\ BAT6 sample (solid blue and red line respectively) have been corrected for absorption (D'Avanzo et al. 2012; Campana et al. 2012) and for dust extinction  (Melandri et al. 2014; Covino et al. 2013), respectively. }
\label{fg1}
\end{figure}
\end{center}

The only difference between OA and GRBs is the orientation of their jets with respect to the line of sight. This allows us to use all the known properties of GRBs detected so far in the $\gamma$--ray band and with well studied afterglow emission to infer the emission characteristics of OA. In particular, OA are normal GRBs with their jet oriented so that \thv$>$\th. 

In order to predict the properties of OA we first need a model describing the entire population of GRBs distributed in the Universe. We use the population code developed recently in \citet[G13 hereafter]{Ghirlanda:2013fq} and extended in \citet[G14 hereafter]{Ghirlanda:2014wt}, called PSYCHE (Population SYnthesis Code and Hydrodynamic Emission model). PSYCHE generates bursts with a redshift $z$ \citep[assigned following the GRB formation rate -][]{Hopkins:2008wo}, a jet opening angle \th, a bulk Lorentz factor \G. The latter two parameters are drawn from two log--normal distributions with median values 5.7$^\circ$ and 90, respectively. These distributions were obtained (G13) in order to reproduce: (a) the \ama\ correlation of a complete (flux limited) sample of \sw\ bursts \citep{Salvaterra:2012si,Nava:2012le}; (b) the flux distribution of GRBs detected by BATSE and GBM/\fe; (c) the detection rate of GRBs by \sw,  \fe\ and \ba. The isotropic equivalent energy \eiso\ and rest frame peak energy \ep\ are obtained, once \th\ and \G\ are extracted, assuming a universal comoving frame, collimation corrected, energetic (see G13 for details)\footnote{ The distribution of \th\ and \G\ are derived self--consistently in G13 to reproduce the $\gamma$-ray properties of GRBs. It is shown in G13 that even changing the assumed values of the comoving frame energetics by a factor of 10 the two inferred distributions are modified accordingly so that the population of GRBs has similar characteristics in terms of energetics and opening angles}. 

Bursts are then assigned a viewing angle \thv\ (according to the $\sin$\thv\ probability distribution) representing the orientation of the jet with respect to the line of sight. Within the GRB population simulated by PSYCHE, there are bursts that can be detected as $\gamma$--ray events because \thv$\le$\th\ and those that that can be detected as orphan afterglows because \thv$>$\th. The latter are the subject of this work. 

In general, since for highly relativistic GRBs seen off--axis only the afterglow can be seen (\S 1), we need to simulate for each burst its afterglow emission. To this aim PSYCHE implements the numerical code \textit{BOXFIT} \cite[VE12 hereafter]{van-Eerten:2011ez,van-Eerten:2012oj}, which is based on numerical 2D simulations of the jet dynamics and assumes synchrotron emission from shock accelerated electrons as the radiation mechanisms of the afterglow phase. The VE12 code considers GRBs with a constant circum burst density. In this standard afterglow model, the external shock emission depends on a set of micro--physical parameters: the index $p$ of the energy distribution of the shock accelerated electrons, the fraction of the dissipated energy distributed to electrons $\epsilon_e$ and to the magnetic field $\epsilon_B$. Finally, the value of the density $n$ for the circumburst medium, is assigned. These are the free parameters determining the afterglow emission of each simulated burst. 

\subsection{Setting the micro--physical shock parameters}

\begin{figure*}
\begin{center}
\includegraphics[width=15cm,trim=20 10 20 20,clip=true]{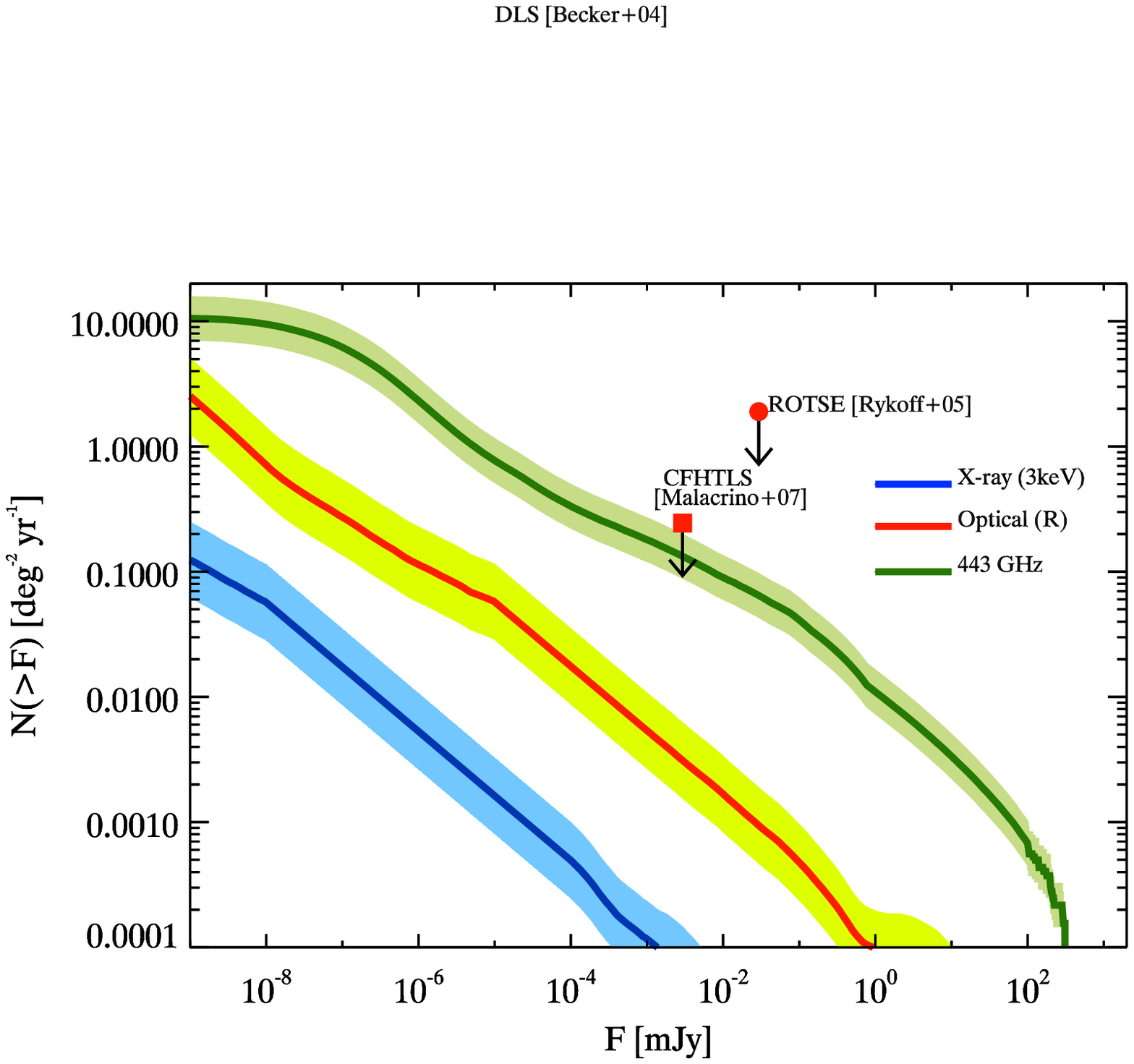}
\caption{Cumulative flux density distribution of orphan afterglows at three characteristic frequencies: {\it R} band for the optical (red line), 3 keV for the X--ray (blue line) and 443 GHz for the mm band (green line - representative of ALMA frequency range). Upper limits of past searches of OA in the optical band are shown by the red symbols (to be compared with the red solid line). The optical $R$ flux distribution is representative of the OA at $z\le4.5$ because, at higher redshift, their $R$ band flux suffers from Ly$\alpha$ suppression. An optical extinction (according to the distribution of Covino et al. 2013) has been applied to the optical fluxes. }
\label{fg2}
\end{center}
\end{figure*}

Distributions of the micro--physical shock parameters ($p$, $\epsilon_e$, $\epsilon_B$) are poorly constrained directly from the observations.  Dense multiwavelength sampling of the afterglow light curve from early times to days after the burst explosion is available for a limited number of bursts. \citet{Panaitescu:2000tx} first derived the values of these parameters through the modelling of the afterglow light curves of 10 GRBs in the pre--\sw\ era. The same pre--\sw\ bursts were used to derive the properties of OA \citep{Totani:2002hb}. However, \sw\ follow up in the X--ray band and optical monitoring campaigns have shown in the last years that the afterglow emission can be very different from burst to burst. The X--ray and optical luminosities at 0.5 days after the burst have proven \citep[e.g.][]{DAvanzo:2012wj,Melandri:2014lk} to be more  dispersed than initially found based on a limited number of events. Moreover, a possible different origin of the X--ray and optical emission, with the latter being more genuinely afterglow, has been proposed \citep{Ghisellini:2009ng}. 

The aim of this work is to estimate the flux level of the population of OA as a whole in order to compare it with the limits of current and future surveys in different bands. To this aim we can use average values of the micro--physical parameters ($p$, $\epsilon_e$, $\epsilon_B$) for the entire simulated GRB population. In order to assign these values, we consider the observed flux of the afterglow of the GRBs composing the complete \sw\ sample \citep[BAT6 -][]{Salvaterra:2012si}. This is a flux limited sample of GRBs and it turns out to have a high level of completeness in redshift (97\%). Therefore, no bias induced by the redshift estimate should be affecting it. Similarly the high flux cut ensures that it is free from any threshold bias related to the GRB detector (BAT on board \sw\ in this case). Fig. \ref{fg1} shows the cumulative distribution of the optical flux (at 12 h after the burst) of the BAT6 sample \citep[red solid line - adapted from][]{Melandri:2014lk}, the cumulative distribution  of the X--ray flux (at 3 keV and at 11h) of the BAT6 \citep[blue solid line - adapted from][]{DAvanzo:2012wj} and the cumulative distribution of the radio flux (at 8.4 GHz between 1 and 6 days) of the BAT6 \citep[green solid line -adapted from][]{Ghirlanda:2013fq}. These are the observed distributions we aim to reproduce assigning to the simulated GRB population a set of values for the micro--physical parameters ($p$, $\epsilon_e$, $\epsilon_B$). 

Since the BAT6 sample contains GRBs detected by \sw\ in the $\gamma$--ray band, we work on the simulated population of GRBs with \thv$\le$\th. We select among them the bright events with the same flux cut adopted for the BAT6, i.e. bursts with a peak flux larger than 2.6 ph cm$^{-2}$ s$^{-1}$ integrated in the 15--150 keV energy band. We assume that the circumburst density $n$ is distributed uniformly between 0.1 and 30 cm$^{-3}$ (also in this case the choice is done in order to reproduce the BAT6 optical and radio flux distributions - Fig. \ref{fg1}) and assign to each simulated burst a value of $n$ randomly extracted from a uniform distribution within this range. We keep the other three parameters ($p, \epsilon_e, \epsilon_B$) fixed. The fiducial values for these three parameters are obtained by reproducing the flux distributions of the BAT6 sample shown by the solid lines in Fig. \ref{fg1} at three different frequencies. With $p=2.3$, $\epsilon_e=0.02$ and $\epsilon_B=0.008$ (as already discussed in Ghirlanda et al. 2013a, 2014) we obtain for the simulated population of GRBs cumulative flux distributions (dashed lines in Fig. \ref{fg1}) which nicely match the radio and optical flux distributions (solid lines in Fig. \ref{fg1}) of the real GRBs of the BAT6 sample. If we assume smaller/larger values for $\epsilon_e$ and $\epsilon_B$ we obtain smaller/larger fluxes in both bands (solid shaded regions in Fig. \ref{fg1}).

Also shown in Fig. \ref{fg1} (solid blue line) is the X--ray flux cumulative distribution at 3 keV computed at 11 h \citep[from][]{DAvanzo:2012wj}. We note that the micro--physical parameter values that reproduce the optical and radio fluxes underestimate the X--ray flux by more than one order of magnitude. This is not unexpected since it has already been discussed in the literature that the X-ray emission of GRBs could be dominated at early times (typically up to half a day after the explosion) by an extra component which is apparently unrelated to the standard afterglow forward shock emission \citep[e.g.][]{Ghisellini:2009ng,DAvanzo:2012wj}. 

Therefore, we assume the microphysical parameters values that reproduce the optical and radio flux distribution of a flux limited sample of real bursts. 


\section{Results}

\begin{table}
\caption{Parameters of the linear fits to the average duration of OA above flux threshold (Fig. \ref{fg3}). Fits parameters of the formula: 
$\log(<T>_{\rm days})=q+m\log(F_{\rm lim,mJy})$}
\label{tab2}
\centering
\begin{tabular}{c c c}
\hline
Band & $m$ & $q$ \\
\hline\hline
X--ray (3 keV) & -0.28 & -1.54 \\
R (7000 $\rm \AA$)	    & -0.36 & -0.72 \\
443 GHz    & -0.44 & 0.44 \\
\hline
\end{tabular}
\end{table}%

\begin{table*}
\caption{Transient surveys in the optical and X--ray bands. On--going and future surveys are marked in boldface. Parameters of the optical surveys, field of view (FOV), cadence, limiting flux $F_{\rm lim}$, coverage and lifetime are from the compilation of \citet{Rau:2009lr}. The rate of orphan afterglow $R_{OA}$ above the survey limiting flux is obtained through the flux density distribution reported in Fig. \ref{fg2}. The average OA duration above this flux limit $\langle T \rangle$ is derived from Fig. \ref{fg3} and from the parameters of the linear fits reported in Tab. \ref{tab2} (the minimum and maximum durations are shown in square parentheses). The last column shows the number of OA per year detectable by the reported surveys. For the X-ray the sky coverage is intended for 24 h. $^{*}$ see http://www.ptf.caltech.edu/ztf  and \citet{Bellm:2014qm}.}
\begin{center}
\begin{tabular}{lllllllll}
\hline
Survey & FOV & Cadence & $F_{lim}$ & Coverage & Lifetime & $R_{OA}$ & $\langle T \rangle$ & \# OA \\
	      & (deg$^2$) & 		&	(mJy)		&	(deg$^2$ night$^{-1}$) & 		days	& (deg$^{-2}$ yr$^{-1}$)	& days & yr$^{-1}$	\\
\hline\hline
PTF	    		& 7.8		& 1m--5d	    	&     1.17$\times 10^{-2}$ 	& 1000		& 			&	1.5$\times10^{-3}$	&	1[0.2-3.8] & 1.5	\\
ROTSE--II	&	3.4		&	1d		&	1.17$\times 10^{-1}$ 	& 450		& 			&	5.2$\times10^{-4}$	&	0.4[0.1-1.7]	 & 0.1 \\
CIDA--QUEST &	5.4		&	2d--1yr	&	4.60$\times 10^{-2}$ 	& 276		&			&	8.0$\times10^{-4}$	&	0.5[0.1-2.3]	&  0.1\\
Palomar--Quest &	9.4		&	0.5h--1d	&	1.17$\times 10^{-2}$ 	& 500		& 2003--2008	&	1.5$\times10^{-3}$	&	1[0.2-3.8] & 0.8	\\
SDSS--II SS	    &	1.5		&	2d		&	2.68$\times 10^{-3}$	& 150		& 2005--2008	&	3.2$\times10^{-3}$	&	1.6[0.4-6.3] & 0.8 	\\
Catilina		    &	2.5		&	10m--1yr	&	4.60$\times 10^{-2}$ 	& 1200		&			&	8.0$\times10^{-4}$		&	0.6[0.1-2.4]	& 0.6 \\
SLS			   &	1.0		&	3d--5yr	&	5.60$\times 10^{-4}$	& 2			& 2003--2008	&	5.2$\times10^{-3}$	&	2.8[0.8-11]	& 0.03 \\	
{\bf SkyMapper}	   &	5.7		&	0.2d--1yr	&	7.39$\times 10^{-2}$ 	& 1000		& 2009--...	&	6.4$\times10^{-4}$	&	0.5[0.2-2.0]	& 0.3 \\	
{\bf Pan--STARRS1}&	7.0		&	3d		&	7.39$\times 10^{-3}$ 	& 6000		& 2009--...	&	2.0$\times10^{-3}$	&	1[0.3-4.4]  & 12	\\	
{\bf LSST}		    &	9.6		&	3d		&	4.66$\times 10^{-4}$	& 3300		& 2022--...	&	5.1$\times10^{-3}$	&	3[0.8-11]	 & 50 \\	
{\bf Gaia}	          & 0.5x2       & 	20d		&    3.00$\times 10^{-2}$		& 2000		&	2014--2019  &    	10$^{-3}$ 		&	1[0.5-5]	& 2\\
{\bf ZTF	$^{*}$}		  & 42.0		& 1d		&	2.00$\times 10^{-2}$       & 22500       & 2017--...           &     1.1$\times 10^{-3}$ &   0.8[0.4-4.8] & 20 \\
\hline
RASS	          & 3.1       &     ...          &     4.00$\times 10^{-5}$      &  12000      & 6 months   &           8.0$\times10^{-4}$	&	1[0.3-4.4]	 & 10 \\	
{\bf eROSITA}         & 0.8     &    6 months & 2.00$\times 10^{-6}$      &  4320$^*$     & 4 years   &           3.0$\times10^{-3}$	&	2[0.5-6.5]	 & 26 \\	
\hline
\end{tabular}
\end{center}
\label{tab1}
\end{table*}%

Following G14 we consider the time when the OA flux reaches its peak (Eq. 3 in G14) and calculate the OA flux at this time of the peak. 
The light curve of the OA will start to rise when $\Gamma\sim1/\sin(\theta_{\rm view}-\theta_{\rm jet})$ and will  peak around the time when all the jet is visible, i.e. 
$\Gamma\sim1/\sin(\theta_{\rm view}+\theta_{\rm jet})$. The OA flux after this time follows the same light curve that an observer would see if with a line of sight within the jet opening angle. 
Therefore, the peak time corresponds to the maximum observed flux from the OA.  According to our simulations, the time when the OA peaks (which depends on the burst parameters and on the viewing angle)  has a broad distribution with a typical value of few hundred days after the GRB (note that this reference is purely theoretical, since the GRB start time is missed in the real case of an OA). 

\begin{figure}
\begin{center}
\includegraphics[width=8.5cm,trim=20 10 20 20,clip=true]{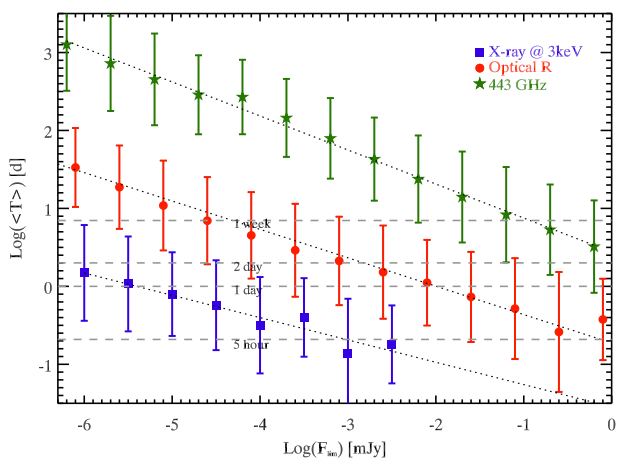}
\caption{Average duration of the simulated population of OA with flux above the corresponding x--axis value. The  bars represent the 1$\sigma$ scatter around the average. Typical timescales are shown by the dashed horizontal lines (as labelled). Linear fits are shown by the dotted lines (fit parameters are reported in Tab. \ref{tab2}. }
\label{fg3}
\end{center}
\end{figure}

\subsection{Orphan afterglows flux density}

The cumulative peak flux density distributions of OA are shown in Fig. \ref{fg2}. The X--ray flux density is computed at 3 keV where the photoelectric absorption by metals in the Galaxy and host galaxy is negligible \citep{Campana:2012sy}. For the optical $R$ flux density, we assumed an $A_{V}$ according to the distribution obtained by the analysis of the BAT6 sample \citep{Covino:2013bl}. This is an asymmetric distribution of $A_{V}$ peaking $<0.5$ mag and extending, in less than 10\% of the cases, to values larger than 1-2 mag.  We note that this assumption is based on the $A_V$ measured in bursts observed within their jet opening angle (i.e. close to the jet axis). Eventually, dust destruction by the GRB X--ray/UV flash could reduce the optical absorption close to the line of sight \citep{Perna:2003uq}. OA, observed at large viewing angles, could therefore have a larger $A_V$ than what we assume. Overall, this effect would further reduce the optical fluxes of OA observed at large viewing angles and consequently reduce the predicted rates of OA detectable in optical surveys.
For the optical we have considered only OA at $z<4.5$ since we expect that the optical emission is fully absorbed by the Ly$\alpha$ absorption at larger redshifts. Also shown in Fig. \ref{fg2} are the range of variation of the flux density distributions obtained by varying the micro--physical parameters (as described in \S2).  

All the flux density distributions shown in Fig. \ref{fg2} represent the flux at the time when the OA reaches its maximum emission. 
On average, OA should reach its maximum flux hundreds of days after the burst. At these times the afterglow emission spectrum peaks at relatively low frequencies (in the mm and radio band). This accounts for the relative normalizations of the flux cumulative distributions at different frequencies in Fig. \ref{fg2}. The three curves, however, converge at very low fluxes (not shown in Fig. \ref{fg2} for clarity) to the total rate of OA which is set by the normalization of the population synthesis code to the rate of GRBs detected in the $\gamma$--ray band by \sw, \ba\ and \fe\ (G13). This can be already seen in the 443 GHz curve of Fig.  \ref{fg2} and corresponds to 10 OA deg$^{-2}$ yr$^{-1}$. 

\subsection{Orphan afterglows timescales}

Surveys can detect orphan afterglows as transient events when the OA flux is above the survey limiting flux. The latter, therefore, determines the rate of detectable OA. However, the survey limiting flux will also determine the OA characteristic duration $\langle T \rangle$. At a fixed frequency, the duration of OA above the survey limiting flux is longer the deeper is the survey. We define $\langle T \rangle$ as the time interval during which the OA flux is larger than the survey flux limit $F_{\rm lim}$. Fig. \ref{fg3} shows the average duration of OA above $F_{\rm lim}$ for the three frequencies considered as reference in this work. 
In general, given the typical flux limits of optical and X--ray surveys (see also \S 4), OA will appear as daily transients. At GHz frequencies, instead they will be much slower transients with duration of even tens--hundreds of days \citep[see also][]{Ghirlanda:2014wt}. Note that this timescale represents the duration of the OA above a given survey limit and it is only due to the instrumental limit. This should not be confused with the timescale of the peak of the OA with respect to the GRB which is due to the combination of the burst geometry (opening angle and viewing angle) and the hydrodynamics (i. e. the deceleration of the fireball, i.e. mainly set by its kinetic energy, initial bulk Lorentz factor and circumburst density). In general, the distribution of the time of the peak of the OA is centred around a few hundreds days. However, at these times the optical emission of OA is extremely faint, so that any conceivable optical survey (also the deepest which will be performed in the future) will detect those OA which peak at relatively early times, i.e. between 1 and 10 days after the trigger, which are the brightest within the population. 

Fig. \ref{fg3} shows that, at any frequency, the OA duration $\langle T \rangle$ increases as the survey limit deepens (i.e. with decreasing survey limiting flux $F_{\rm lim}$). Tab. \ref{tab2} shows the parameters (slope $m$ and normalization $q$) of the linear fit (dotted lines in Fig. \ref{fg3}) to the data shown in Fig. \ref{fg3} for the three characteristic frequencies. 

\section{Orphan afterglows detection rate}

In this section we compare our results with past searches for OA in the optical and X--ray band and show specific predictions for on--going or planned surveys in these bands.  We also consider forthcoming large projects like the Large Synoptic Sky Telescope (LSST)  and the extended ROentgen Survey with the Imaging Telescope Array (eROSITA)  which will conduct almost all sky surveys in the optical and X--ray bands, respectively. For the radio band, \citet{Ghirlanda:2014wt} showed that the OA rates are consistent with the (upper) limits of past radio surveys which did not detect any credible orphan afterglow. Forthcoming radio surveys like the VAST/ASKAP at 1.4 GHz or the MeerKAT or EVLA at 8.4 GHz  could detect 3$\times10^{-3}$ and  3$\times10^{-1}$ OA deg$^{-2}$ yr$^{-1}$, respectively.  The deeper SKA survey, reaching the $\mu$Jy flux limit, could detect up to 0.2--1.5 OA deg$^{-2}$ yr$^{-1}$ \citep{Ghirlanda:2014wt}. Here we report the predictions for the optical and X--ray surveys.

\subsection{Optical surveys} 

Among past searches for orphan afterglows in the optical, \citet{Rykoff:2005oe} used the Robotic Optical Transients Search Experiment III (ROTSE--III). Over a period of 1.5 year, they identified no credible GRB afterglow. They place a 95\% upper limit on the OA rate of 1.9 deg$^{-2}$ yr$^{-1}$ at $R=20$. The Deep Lensing Survey \citep[DLS -][]{Becker:2004bx} provides a (less constraining) limit of 5.2  deg$^{-2}$ day$^{-1}$ for transients with typical duration of a few ksec and 19.5$<R<$23.4. \citet{Malacrino:2007ud} obtained a more stringent upper limit from the CFHTLS Very Wide survey: excluding that the three transient they find are GRBs \citep{Malacrino:2007ud}, an upper limit of 0.24 deg$^{-2}$ yr$^{-1}$ down to $R=23$ can be placed. The ROTSE--III and CFHTLS limits are shown in Fig. \ref{fg2} (filled red symbols) and they are consistent with the rate for the optical band predicted by our model (solid red line in Fig. \ref{fg2}). Also, no credible OA was found in the Faint Sky Variability Survey project \citep{vreeswijk:2002qy}.

The present and future major surveys in the optical are shown in Tab.~\ref{tab1}. Most of the optical survey parameters are obtained from \citet{Rau:2009lr}. In Tab. \ref{tab1} we report the survey name (Col.1), its field of view (FOV) and its cadence (Col. 2 and 3). The limiting flux density and the coverage, representing the sky area covered per night, are reported in Col. 3 and Col. 4, respectively. Through Fig. \ref{fg2} we can derive the rate $R_{OA}$ of OA that have their peak flux density above each survey limiting flux. This is reported in units of deg$^{-2}$ yr$^{-1}$ in Col. 7 of Tab. \ref{tab1}. Similarly, from Fig. \ref{fg3} it is possible to estimate the average duration $\langle T \rangle$ of the OA above the survey limiting flux (Col. 8 in Tab. \ref{tab1}). In square parentheses we indicate the upper and lower estimates of the average duration (i.e. corresponding to the 1$\sigma$ error bars in Fig. \ref{fg3}). We derive the rate of OA (expressed in number of OA per year - last column of Tab. \ref{tab1}) in a given survey as $N_{OA} = R_{OA}\times C\times \langle T\rangle$, where $C$ is the fraction of the sky covered by the survey per night (coverage in Tab. \ref{tab1})

\subsection{X--ray surveys} 

Searching for GRB afterglows in X--ray surveys led to the discovery of few flare stars \citep{Grindlay:1999yf,Greiner:2000zp}. The 27 X--ray transients, detected in the 5.5 year survey of Ariel V \citep{Pye:1983qc}, provide a conservative upper limit of $1.15\times 10^{-3}$ deg$^{-2}$ yr$^{-1}$ corresponding to a flux $\approx$0.06 mJy \citep{Grindlay:1999yf}. This is consistent with our predictions for the X--ray band (solid blue line in Fig. \ref{fg2}) at the same flux limit. 


Among the widest X--ray surveys, the ROSAT All--Sky Survey covered the full sky reaching a limiting flux of 4$\times 10^{-13}$ erg cm$^{-2}$ s$^{-1}$ (0.5--2 keV) in almost 6 months. This flux limit (assuming a spectrum with photon index -2) corresponds to a flux density of $\sim 4\times 10^{-5}$ mJy which, according to our estimates, gives a rate $R_{OA}\sim 8\times10^{-4}$ deg$^{-2}$ yr$^{-1}$. The RASS scan procedure covered a full sky circle of width 2 deg every orbit corresponding to $\sim$ 12000 deg$^2$ day$^{-1}$. According to our estimates (Fig. \ref{fg2} and Tab. \ref{tab1}), the typical duration of the OA above the RASS flux limit should be of $\sim$ 1 day so that the expected OA number should be $\sim$ 4.8 during the 6 months survey lifetime. This result is consistent with the estimate of \citet{Greiner:2000zp}. They effectively searched in the RASS for GRB afterglows and concluded that of the 23 candidates only a few could be due to GRBs, most of the others being flare stars. The second release of the RASS, 2RXS\footnote{http://xmm.esac.esa.int/external/xmm\_science/workshops/\\2014symposium} has been extended to a flux limit a factor of 4 deeper than the first release. Therefore, we expect to have $\sim$12 OA in the 2RXS.

In the X--ray band {\it Chandra} and {\it XMM-Newton} have performed deep surveys but, due to their small field of view, at the expense of a relatively small portion of the sky explored \citep[see][]{Brandt:2005qp}. The observing strategy in these surveys was not a scanning mode as the RASS, but rather the combination of pointed repeated observations of the same field. Therefore, it is difficult to reconstruct the overall sky coverage. As a gross estimate we can compute the expected number of OA by multiplying the predicted rate (according to our results of Fig. \ref{fg2}) times the area of the sky covered. We stress that this is an overestimate of the number of OA that could be detected by these surveys. Among the deepest surveys,  the 2Ms {\it Chandra} Deep Field North covered 0.13 deg$^2$ in the 0.5-8.0 keV band down to a flux limit of $\sim 10^{-16}$ erg cm$^{-2}$ s$^{-1}$ \citep{Alexander:2003tk}. To such a flux limit we predict less than 10$^{-2}$ OA yr$^{-1}$. Similar rates are expected in the {\it XMM--Newton} Large Scale Survey \citep{Pierre:2004hh} which, with a sensitivity of $\sim 5\times10^{-15}$ erg cm$^{-2}$ s$^{-1}$ (0.5--2 keV) and a 10 deg$^2$ of sky coverage, should detect at most 0.1 OA yr$^{-1}$. Both {\it Chandra} and {\it XMM-Newton} have performed several other surveys, but, despite with larger sky coverage than those mentioned above, at the expense of their sensitivity (e.g. the {\it XMM--Newton} Bright Serendipitous Survey - \citet{Della-Ceca:2004qr}). According to our predictions, therefore, there seems to be no chance to have detected any orphan afterglow in current deep X--ray surveys\footnote{In the XMM/EPIC database, whose variability richness will be
fully explored by the EXTraS project \citep{De-Luca:2015lr}, we do not expect orphan afterglows to be present.}.


\subsection{Orphan Afterglow distinguishing properties}


The most important question to address is how to distinguish OA from other transients when they will be detected in large sky surveys. As shown in Fig. \ref{fg3} OA will appear as daily transients in optical and X--ray surveys, given the typical flux limits of current and forthcoming surveys (see Tab. \ref{tab1}), and many other extragalactic sources will have similar duration. They will show a decaying light curve with a temporal slope which is that of typical GRB afterglows, i.e. $\propto t^{-\delta}$ with $\delta\sim 1-2$ but not uniquely characterising GRBs as a class. The lack of any associated high energy $\gamma$--ray counterpart will hamper their classification as orphan afterglows of GRBs. The success of a transient survey is that of classifying, after discovery, the detected transients and to this aim a dedicated follow up program is fundamental. 

A first way to assess the orphan afterglow nature of the transient is by a systematic optical photometric and spectroscopic follow-up. The optical/X-ray light-curves, especially if the OA that is detected is still quite bright and/or before its light-curve peak, can be a very useful tool for a preliminary source classification. Indeed, its shape and its decay power-law index should be different from those of SNe or blazars. In addition to the ground-based optical facilities, the future satellite SVOM \citep{Basa:2008fk}, to be operational at the same epoch of LSST, will be able to perform simultaneous X-ray and optical observations of OA candidates. 

A final identification is given by optical spectroscopy. The spectral continuum and the absorption lines present in the optical-NIR spectra of the  afterglow are very different from those typically found in SNe or blazars \citep{Fynbo:2009kx,Christensen:2011fj}. Nonetheless, spectra with a sufficient signal-to-noise ratio are needed. To date this requirement is fulfilled down to $R \sim$22 with a reasonable amount of integration time ($\sim$2hrs) (e.g.: with X-shooter at ESO/VLT). Future larger telescope will make possible to obtain similar results also for fainter objects. Therefore, with the possible increase of transient detections from future surveys, a considerable amount of telescope time may be necessary for a systematic optical follow-up of unknown transients and of potential OA candidates. Among  projects which could substantially contribute to the broad band (optical to NIR) spectroscopy of transients discovered in surveys there is the Son Of X--Shooter (SOXS - P.I. Campana) which has been proposed for the NTT with a considerable number of dedicated nights per year  and the Nordic Optical Telescope Transient Explorer (NTE)\footnote{http://dark.nbi.ku.dk/news/2014/nte\_is\_a\_go/}.  Another way to discriminate between OA and other transients comes from the analysis of the broad Spectral Energy Distribution (SED). Here we compare the typical SED of OA with that of potential competitors extragalactic sources like SNe and blazars. 

\begin{figure*}
\begin{center}
\includegraphics[width=15.cm,trim=20 10 15 20,clip=true]{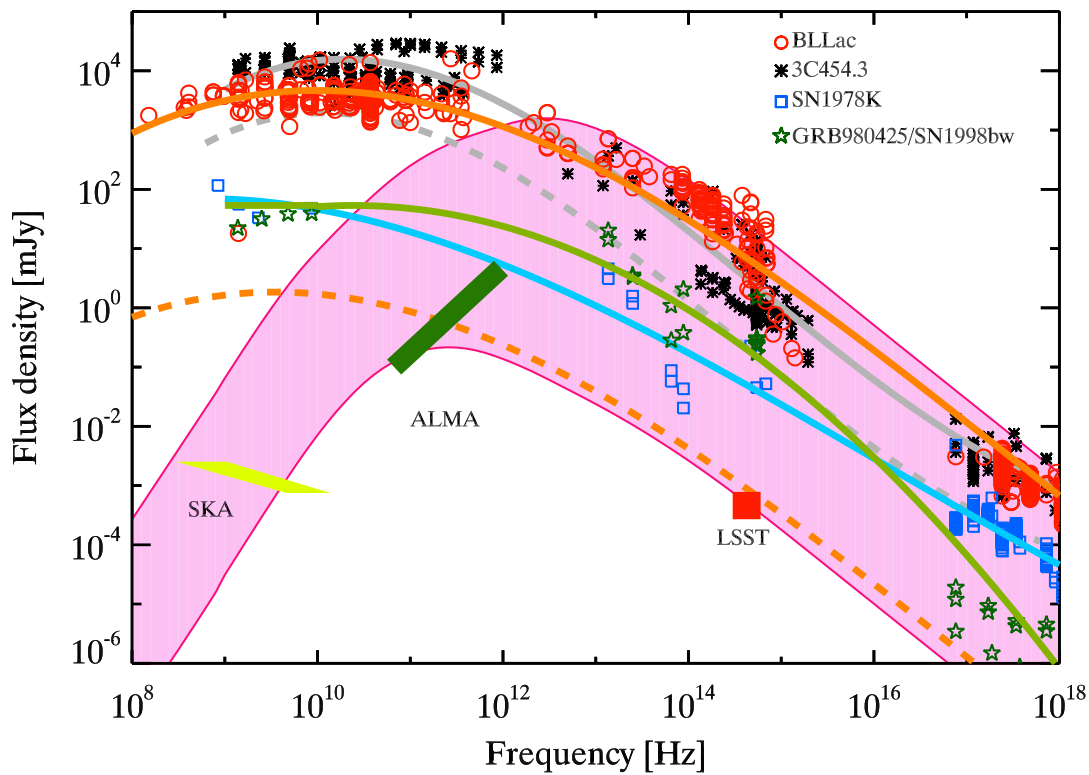}
\caption{Spectral energy distribution of the OA that can be detected by the LSST (pink filled region). The SED of the low power blazar BLLac (open circles), of the FSRQ 3C454.3 (asterisks) and of two supernovae SN1978K (open squares) and the GRB--SN associated GRB980425/SN1998bw (open stars) are shown by different symbols. The solid lines provide an interpolation of the data points and do not represent any physical model. For the two blazars we also show (dashed grey line for 3C454.3 and dashed orange line for BLLac) how their spectra would appear if they were at $z=2$ (typical of GRBs). The (5$\sigma$) limits for a 12 h continuum observation with the SKA is shown by the yellow shaded region. The green shaded region marks the limiting flux of an ALMA observation (32 antennas of 12 m for 3 h of observation in dual mode - from https://almascience.eso.org/proposing/sensitivity-calculator). The LSST liming flux (see Tab.\ref{tab1}) is shown by the red square symbol. }
\label{fg4}
\end{center}
\end{figure*}

Since the most promising detections will be with the forthcoming LSST (Ivezic et al. 2008 - see Tab. \ref{tab1}), we have considered only the OA that will be detected by this survey. We predict a rate $\sim$ 50 OA per year. The overall SED (i.e. the convolution of the SEDs of all OA detectable by the LSST survey) is shown by the hatched pink region in Fig. \ref{fg4}. The typical SED of OA detectable by LSST peaks in the 10$^{11-13}$ Hz range. The spectrum below the peak, in the GHz down to the MHz range scales $\propto \nu^2$. 

Possible extragalactic variable sources that could compete with GRB orphan afterglows in brightness, frequency of discovery and timescales are supernovae and blazars. Fig. \ref{fg4} shows the SED of two blazars: the Flat Spectrum Radio Quasar (FSRQ) 3C454.3 and BLLac itself as representative of the respective classes \citep[see][]{Ghisellini:2010wf}. Two supernovae are also shown: SN 1978K as a possible representative of highly luminous supernovae and the GRB980425/SN1998bw \citep{Galama:1998ca} for the class of GRB--SNe associated. For all these sources, we report their SED as obtained by multifrequency observational campaigns and retrieved from Italian Space Agency (ASI) Science Data Center Sed Builder tool\footnote{http://tools.asdc.asi.it}. The solid curves in Fig. \ref{fg4} are not physical models but only illustrative of the overall broad band spectral energy distribution of these classes of objects. For the blazars we also show how their SED would be like if they were shifted at $z=2$, i.e. at the typical distance of long GRBs. 

For comparison in Fig. \ref{fg4} we show the LSST flux limit (red square symbol). The OA that can be detected by LSST when their jet emission is fully visible by the off--axis observer will have their peak frequency already below the optical band, in the mm region. This is because the peak of the OA emission is reached several months after the burst (\S 3 - see also G14). Furthermore, Fig. \ref{fg4} shows that their emission in the MHz/GHz region is still in the self absorbed regime. Differently, BLLacs and SN emission like 1998bw or 1978K are characterised by a softer spectrum in the radio band than the typical OA detected in an optical survey like the LSST. Therefore, the follow up of these transients in the mm and GHz bands will characterise their different SED. 

\section{Discussion}

Among previous works in the literature which estimated the detection rate of OA, \citet{Totani:2002hb} considered 10 GRB of the pre--\sw\ era with well monitored afterglow light curves as templates. By assuming different off--axis viewing angles they estimated the rate of OA in the X--ray, optical and radio band. Their predictions were based on a very small number of afterglows mostly representative of the bright afterglow population of GRBs. Similarly \citet{Zou:2007uk} adopted a set of fixed physical parameters (kinetic energy and micro--physical parameters) and allowed only for a possible distribution of \th. They predict a rate of 1.3$\times 10^{-2}$ deg$^{-2}$ yr$^{-1}$ for OA brighter than $R=20$ which is consistent with our findings at the corresponding flux (red solid curve in Fig. \ref{fg2}). However, their flux distributions appear steeper than our model at low fluxes, thus predicting a larger rate of OA in surveys that go deeper than the above limit. We consider that our estimates better represent the low flux end of the OA distribution since our code is calibrated with the entire GRB prompt emission flux distribution and includes a more representative sample of afterglows to fix the micro--physical parameters. 

What is new in our model is that we predict the properties of OA based on the observed properties of GRBs in the $\gamma$--ray band considering as constraints the flux and fluence distribution of the population of GRBs detected by \sw, \ba\ and \fe. Since the $\gamma$--ray energy detected in the prompt emission is a proxy of the kinetic energy driving the afterglow deceleration, our simulated population of bursts includes both high and low kinetic power bursts. The choice to reproduce the afterglow flux distribution of the complete BAT6 sample of \sw\ bursts, despite being composed by relatively bright events, ensures that we are extending the flux distribution of the synthetic GRB population to the low end better than what could be done with the limited number of GRBs detected in the pre--\sw\ era. 

We have assumed that GRBs have a jet with a top--hat uniform structure, i.e. the kinetic energy and the bulk Lorentz factor are constant within the jet opening angle. Alternatively \citep{Rossi:2002ys,Zhang:2002sy} GRB jets could be structured, i.e.  the kinetic energy (and possibly also) the bulk Lorentz factor depend (as a power--law or exponentially) on the angle from the jet axis. In the former scenario, considered in this work, orphan afterglows are naturally expected to dominate the number of GRBs in the Universe (considering a typical jet opening angle of few degrees). Even if the top--hat uniform jet emission can be seen when \thv$\ge$\th, its flux decreases drastically for off--axis observers so that we can consider valid the approximation so far understated that we do see the prompt emission only of GRBs whose uniform jet is pointed towards the Earth (i.e. when \thv$\le$\th). In the structured jet model, instead, there is always a portion of the jet that is pointing to the observer. Therefore, the observed GRB properties depend only on the viewing angle \thv\ so that orphan afterglows should not exist in principle, since even at large angles from the jet there is jet emission that can be seen \citep{Salafia:2015oq}. However, also in this scenario OA could still be present if either the prompt emission at large angles is below any detector threshold or if the jet is uniform within a relatively narrow core and highly structured (i.e. with a steeply decreasing energy profile) outside it, as suggested by recent results from the modelling of the luminosity function \citep{Pescalli:2015cj}. In this model the detection rates of OA would be, anyway, smaller than in the top--hat model adopted here as shown in \cite{Rossi:2008om}. 

Finally, note that in Fig. \ref{fg1} we have shown that our afterglow model does not reproduce the X--ray early flux observed in the BAT6 sample of GRBs. This is due to the fact that at early times the X--ray emission has been shown to be inconsistent with (i.e. larger than) the forward external shock model \citep[e.g.][]{Ghisellini:2009ng}. However, we have used the afterglow emission of the population of OA (as shown in Fig. \ref{fg2}) in predicting the rate of detection of OA by forthcoming X--ray surveys. Indeed, the time when the brightest OA peak in the X--ray (i.e. those that will likely be above the threshold of forthcoming X--ray surveys) is a few days after the prompt emission i.e. when also in the X--ray the emission turn out to be dominated by the afterglow component \citep{Ghisellini:2009ng,DAvanzo:2012wj}. 

\section{Conclusions}

We computed the emission properties of the population of orphan afterglows in the optical and X--ray band. Our simulation procedure relies on what we have so far observed for a well defined and complete sample of GRBs detected by \sw\ \citep{Salvaterra:2012si} for which the X--ray, optical and radio emission have been extensively studied \citep{Campana:2012sy,DAvanzo:2012wj,Covino:2013bl,Melandri:2014lk,Ghirlanda:2013bq}. 

Our model allows us to predict the expected rate of detection of OA in past, current and future optical and X--ray surveys (Tab. \ref{tab1}). For a similar work for the radio band see \citet{Ghirlanda:2014wt} (see also \citet{Metzger:2015eq}). Most past and on--going optical surveys have small chances to detect OA. Among these, the Palomar Transient Factory (PTF - \citet{Law:2009wu}) could marginally see one OA per year  \citep[consistent with][predictions]{Rau:2009lr}  given its relatively low sensitivity compensated by the large portion of the sky covered per night (10$^3$ deg$^2$). Instead, according to our model, an optical survey like that of Pan--STARRS1 which will cover 6000 deg$^2$ per night could already detect a dozen of OA per year. Larger detection rates are expected for the forthcoming development of the PTF survey. The Zwicky Transient Facility \citep{Bellm:2014qm}, which is designed specifically for transients discovery, will cover about 22500 deg$^{-2}$ per night down to a limiting magnitude $\sim$20.5. We expect that it will detect $\sim$20 OA yr$^{-1}$. A considerably larger number of  OA will be accessible by the Large Synoptic Sky Telescope survey \citep[LSST --][]{Ivezic:2008qd}. The telescope will have a 9.6 deg$^{2}$ field of view and will be able to survey 10$^{4}$ deg$^{2}$ of the sky every three nights down to a limiting magnitude for point sources $R\sim$24.5. With these parameters we estimate it could detect 50 OA yr$^{-1}$. 

An interesting prediction concerns the {\it Gaia} satellite \citep{Lindegren:2010fb}. It will carry two telescopes each one with a field of view of 0.7$^\circ$x0.7$^\circ$ and will scan an angle of 360$^\circ$ every six hours. Therefore, it will cover $\sim$2000 deg$^2$ per day performing a survey down to a limiting flux of 0.03 mJy. According to our model $R_{OA}\sim10^{-3}$ deg$^{-2}$ yr$^{-1}$ at this flux limit so that we predict that {\it Gaia} will detect about 10--15 OA in its 5 year mission. This estimate is consistent with that reported in \citet{Japelj:2011df}.

Given the depth of forthcoming optical surveys we expect that OA will have a typical redshift $z\sim$2. At such distances the typical GRB host galaxy should be fainter than the LSST limiting magnitude \citep{Hjorth:2012xu}.

The difficulty will be to disentangle these OA from other (galactic and extragalactic) transients that will be detected with similar flux and temporal behaviour. The follow up in the optical and X--ray will secure the sampling of the light curve which could be the first hint to the OA nature (with respect to potential other extragalactic transients like supernovae and blazars).   The availability of dedicated facilities or assigned observing time at different ground based facilities will be crucial in this respect. Optical/NIR spectroscopy will discriminate extragalactic transients \citep[e.g.][]{Gorosabel:2002yq}, low frequency (mm and radio bands) observations\footnote{The LSST will start operating approximately in the same period of the Square Kilometer Array (SKA).}  could be used (as shown in Fig. \ref{fg4}) to distinguish among possible competing transients sources. For particularly low redshift transients, the search for the host galaxy could also provide further clues on their nature. 

Here we have computed the OA flux cumulative distribution at the reference frequency of 443 GHz which is one of the frequencies covered by e.g. ALMA. 
We have verified that the few hundred GHz range is where OA are brightest considering the typical timescales when they become visible (\S 2). Indeed, at 40 GHz and 4000 GHz the flux cumulative curves in Fig. \ref{fg2} lie below the solid green curve representing the flux density at 443 GHz.
The Herschel/SPIRE survey ATLAS\footnote{http://www.h-atlas.org/ \newline http://www.h-atlas.org/survey/fields}, one of the widest covering a total of 500 deg$^2$, is limited by the confusion limit of 5--7 mJy at 250--500 $\mu$m so that we expect less than 0.1 OA yr$^{-1}$. Spitzer SWIRE\footnote{http://swire.ipac.caltech.edu/swire/astronomers/program.html\newline http://swire.ipac.caltech.edu/swire/public/faqs.html\#where} observed six fields in the northern and southern sky with typical areas between $\sim$4.2 deg$^2$ and $\sim$12 deg$^2$ with higher sensitivities of few tens of $\mu$Jy in the low frequencies channels (IRAC) at 3.6$\mu$m and 4.5$\mu$m. These fields were covered on different timescales between 1 and 6 days. According to our model we expect a rate of less than one OA per year in such fields above the deepest flux limits of this survey.

Among forthcoming X--ray surveys we consider the extended ROentgen Survey with the Imaging Telescope Array (eROSITA - Merloni et al. 2012) which will cover the full sky up to 10 keV with a flux limit $\sim 2\times10^{-14}$ erg$^{-1}$ cm$^{-2}$ s$^{-1}$ in the 0.5--2 keV band. Therefore, $\sim 3\times10^{-3}$ deg$^{-2}$ yr$^{-1}$ OA should be reachable by this survey (Tab. \ref{tab1}). According to the planned scanning strategy,  a full circle of width 2 degree will be covered every four hours. This corresponds to $\sim$ 4320 deg$^2$ day$^{-1}$. The expected OA number is $\sim$26 yr$^{-1}$ \citep[but see also][]{Khabibullin:2012lr}. A larger number of OA (by a factor 2) could be reached by the WFXT survey \citep[e.g.][]{Rosati:2011ft}.


\section*{Acknowledgments}
ASI I/004/11/0 and the 2011 PRIN-INAF grant are acknowledged for financial support. S. Covino, G. Ghisellini and L. Nava are acknowledged for stimulating discussions. Development of the Boxfit code (van Eerten et al. 2012) was supported in part by NASA through grant NNX10AF62G issued through the Astrophysics Theory Program and by the NSF through grant AST-1009863.  DB is funded through ARC grant DP110102034.  We acknowledge the anonymous referee for useful comments that helped us to improve the manuscript. 

\bibliographystyle{aa}
\bibliography{bibliografia}

\end{document}